\documentclass[10pt]{article}
\usepackage[super,compress]{cite}

\usepackage{fancyhdr}
\usepackage{color}
\usepackage{amsmath}
\usepackage{amsfonts}
\usepackage{latexsym}
\usepackage{epsfig}

\usepackage{graphicx}
\usepackage{verbatim}
\usepackage{psfrag}
\usepackage{multirow}
\usepackage{ragged2e}
\usepackage{cancel}
\usepackage{color}
\usepackage{xcolor}
\usepackage{hyperref}

\date{}

\begin{document}



%
%
\newcommand{\ba}{\begin{eqnarray}}
\newcommand{\ea}{\end{eqnarray}}
\newcommand{\be}{\begin{equation}}
\newcommand{\ee}{\end{equation}}
\newcommand{\bd}{\begin{displaymath}}
\newcommand{\ed}{\end{displaymath}}
\newcommand{\bn}{\begin{enumerate}}
\newcommand{\en}{\end{enumerate}}
\newcommand{\pa}{\partial}
\newcommand{\f}{\frac}
\newcommand{\bp}{\begin{pmatrix}}
\newcommand{\ep}{\end{pmatrix}}
\newcommand{\ci}{\cite}
\newcommand{\eps}{\epsilon}
\newcommand{\heta}{\hat \eta}
\newcommand{\del}{\delta}
\newtheorem{truth}{Theorem}
\newtheorem{prob}{Problem}
\newtheorem{corl}{Corollary}
\newtheorem{rem}{Remark}
\newcommand{\hf}{\frac12}
\newcommand{\vO}{\vec \Omega}
\newcommand{\hu}{\hat u}
\newcommand{\dud}[2]{\frac{\partial {#1}}{\partial{#2}}}
\newcommand{\ddud}[2]{\frac{\partial^2{#1}}{\partial{#2^2}}}
\newcommand{\lfp}[2]{L_{FP}(j_{#1},j_{#2})}
\definecolor{darkgreen}{rgb}{0.,0.6,0.2}
\newcommand{\vtr}[3]{
    \begin{pmatrix}
    {#1} \\
    {#2} \\
    {#3}
    \end{pmatrix}
}
\newcommand{\mub}[1]{\underline{\mathbf{{#1}}}}
\newcommand{\ve}{\mub \eta}
\definecolor{darkgreen}{rgb}{0.,0.6,0.2}

\title{The Bloch equation for
spin dynamics in electron storage rings:
computational and theoretical aspects {\thanks{
Based on a talk at ICAP18, Key West, Florida, USA, Oct 19--23, 2018 and on 
Ref. \citen{SDE}. Also available as DESY Report 19-156.}}}

\maketitle


\centerline{Klaus Heinemann \footnote{Corresponding author.} }
\centerline{\em Department of Mathematics and Statistics, University of New Mexico,} 
\centerline{\em Albuquerque, NM 87131, USA} 
\centerline{\em heineman@math.unm.edu}   
\vspace{4mm}
\centerline{Daniel Appel\"o }  
\centerline{\em Department of Applied Mathematics, University of Colorado Boulder,} 
\centerline{\em Boulder, CO 80309-0526, USA } 
\centerline{\em Daniel.Appelo@Colorado.edu}  

\vspace{4mm}
\centerline{\em Desmond P. Barber} 
\centerline{\em Deutsches Elektronen-Synchrotron (DESY) }
\centerline{\em Hamburg, 22607, Germany}
\centerline{\em and:}
\centerline{\em Department of Mathematics and Statistics, University of New Mexico}
\centerline{\em Albuquerque, NM 87131, USA}
\centerline{\em mpybar@mail.desy.de}

\vspace{4mm}
\centerline{\em Oleksii Beznosov}
\centerline{\em Department of Mathematics and Statistics, University of New Mexico,} 
\centerline{\em Albuquerque, NM 87131, USA}
\centerline{\em dohter@protonmail.com}

\vspace{4mm}
\centerline{\em James A. Ellison}
\centerline{\em Department of Mathematics and Statistics, University of New Mexico,} 
\centerline{\em Albuquerque, NM 87131, USA}
\centerline{\em ellison@math.unm.edu}



\vspace{6mm}

\newpage

\begin{abstract}
In this paper we describe our work on spin polarization in high-energy electron storage rings which we
base on the Bloch equation for the polarization density and which aims towards the $e^{-}-e^{+}$ option of
the proposed Future Circular Collider (FCC-ee) and the proposed Circular Electron Positron Collider
(CEPC). The  Bloch equation takes into account non spin-flip and spin-flip effects due to
synchrotron radiation including the spin-diffusion effects and the Sokolov-Ternov effect with its
Baier-Katkov generalization as well as the kinetic-polarization effect. 
This mathematical model is an alternative
to the standard mathematical model 
based on the Derbenev-Kondratenko formulas. For our numerical and analytical
studies of the Bloch equation we develop an approximation to the latter to obtain an effective 
Bloch equation. This is accomplished by finding a 
third mathematical model based on a
system of stochastic differential equations underlying
the Bloch equation and by approximating that system via the method of averaging from perturbative
ODE theory. We also give an overview of our algorithm for numerically integrating the effective 
Bloch equation. This discretizes the phase space using spectral methods and discretizes time via the
additive Runge-Kutta method which is a high-order semi-implicit method. 
We also discuss the relevance of the third mathematical model for
spin tracking.

\vspace{5mm}

{\tiny{.}} \hspace{12mm} differential equations, method of averaging. 
\end{abstract}


\newpage

\tableofcontents

\newpage

\section{Introduction}

In this paper we describe some analytical and numerical aspects
of our work on spin polarization in high-energy electron storage
rings aimed towards the  $e^{-}-e^{+}$ option of
the proposed Future Circular Collider (FCC-ee)
and the proposed Circular Electron Positron Collider (CEPC).
The main questions for high-energy rings like
the FCC-ee and CEPC are: (i)
Can one get polarization? (ii) What
are the theoretical limits of the  polarization?
This paper builds on our ICAP18 papers and talks \cite{HABBE,OB}, as well as 
a talk at an 
IAS mini-workshop on Beam Polarization.\cite{IAS}

Photon emission in synchrotron radiation affects the orbital motion of
electrons in a storage ring and can lead to an equilibrium bunch
density in phase space. This is modeled by adding noise and damping to
the particle motion\cite{Sa,CT2}. The photon emission also affects the
spin motion and this can lead to the build-up of spin
polarization which can reach an equilibrium resulting from a balance
of three factors, namely the so-called Sokolov-Ternov process, driving
build-up, depolarization and the so-called kinetic polarization
effect.

The Sokolov-Ternov process \cite{st64} causes a build up of the
polarization due to an asymmetry in the spin-flip transition rates for spin up and spin down along a spin-quantization axis. The roots here are
in the Dirac equation.
The depolarization can be viewed as a consequence of the trajectory
noise feeding through to the spin motion via the spin-orbit coupling
embodied in the Thomas-BMT equation \cite{jackbook} and thus leading to spin
diffusion. 
The kinetic polarization is also a result of spin-orbit coupling. 

The three factors have been modeled mathematically in two ways, the first based on Ref. \citen{DK73} by Derbenev and Kondratenko (see also Ref. \citen{Mane}) and the second
on Ref. \citen{DK75}, also by Derbenev and Kondratenko.
Here we discuss the second model and then introduce a new, third, mathematical model, based on stochastic differential equations (SDEs).



So far, analytical estimates of the attainable polarization have been based on
the aforementioned Ref. \citen{DK73} via the
so-called {\it Derbenev-Kondratenko formulas} \cite{Barber-Handbook}. 
A recent overview is part of Ref. \citen{IAS}. In analogy with studies of the trajectories of single particles, this model leans towards the study of single spins and it relies in part on plausible assumptions grounded in deep physical intuition with the introduction of a field of
spin-quantization axes, the so-called {\it invariant spin field} (ISF), erected on the six-dimensional phase space \cite{BEH}. Here, the depolarization and the kinetic polarization
follow from the geometry of the ISF.

For the future, a third question for high-energy rings like the FCC-ee and CEPC is: are the Derbenev-Kondratenko formulas complete? 
We believe that the model based on the Derbenev-Kondratenko formulas is an approximation of the model from Ref. \citen{DK75} mentioned above which is
based on the so-called {\it polarization density} of the bunch.
In this model one studies the evolution of the bunch density in phase space with the Fokker-Planck equation (\ref{eq:n10a}).
The corresponding equation for spin is the evolution equation (\ref{eq:n10})
for the polarization density which we call the  Bloch
equation (BE) and which generalizes the orbital Fokker-Planck equation
\footnote{Note that in Ref. \citen{SDE} we use the term ``Full Bloch equation''
instead of simply ``Bloch equation".}.
We use the name ``Bloch'' to reflect the analogy with equations
for magnetization in condensed matter \cite{Bloch}.
Each of the above three synchrotron-radiation effects corresponds to terms in the BE. Thus it takes into account effects on spin due to synchrotron radiation including the spin-diffusion effects, the Sokolov-Ternov effect with its Baier-Katkov generalization, as well as the kinetic-polarization effect.

The BE was introduced by Derbenev and Kondratenko in 1975 \cite{DK75} as a generalization to the whole phase space (with its noisy trajectories) of the Baier- Katkov-Strakhovenko (BKS) equation which just describes the evolution of polarization by spin flip along a single deterministic trajectory \cite{BKS,IAS}. The BE is a system of three Fokker-Planck-like equations for the three components of the polarization density coupled by a Thomas-BMT term and the BKS terms but uncoupled within the Fokker-Planck terms. The integral of the polarization density is the polarization vector of the bunch. We remark that the polarization density is proportional to the phase space density of the spin angular momentum. See Refs. \citen{IAS} and \citen{Eliana} for recent reviews of polarization history and phenomenology.
Thus, we study the initial-value problem of the system of coupled orbital 
Fokker-Planck equation and the BE.
The third model is based on the system of coupled spin-orbit
SDEs (\ref{eq:4.4.18e}) and (\ref{eq:4.4.18c}) 
and its associated Fokker-Planck equation which governs the evolution of the (joint) spin-orbit probability density. The third model is 
equivalent to the second model, i.e., the one based on Ref. 
\citen{DK75},
but we believe that the third model is also more amenable to analysis.

We proceed as follows.
In the second section we present the BE for the
laboratory frame. We also introduce our newly discovered 
system of stochastic differential equations (SDEs) which underlie
the whole BE.
Thus we can 
model the BE in terms of white noise in the SDEs, thereby
extending the classical treatment of spin diffusion 
from Ref. \citen{BH} to 
a classical treatment of all terms of the BE. So
we have extended the classical model of 
spin diffusion to a classical model which includes 
the Sokolov-Ternov effect, its Baier-Katkov correction, and
the kinetic-polarization effect.
As an aside this may lead to a new 
Monte-Carlo approach to simulation which includes these effects, using
modern techniques for integrating SDEs.
The second section also presents the reduced Bloch equation (RBE)
obtained by neglecting
the spin flip terms and the kinetic-polarization term in the BE.
The RBE approximation is sufficient for computing the physically interesting 
depolarization time and it shares the terms with the BE that are most challenging to discretize. 
Thus in this paper, when we consider the discretization,
we only do it for the RBE.
In the third section we discuss the RBE in the beam frame and the underlying
stochastic differential equations.
In Section  4 we derive an effective RBE by applying
the method of averaging to the underlying
stochastic differential equations.
In the fifth section we outline
our algorithm for integrating the effective RBE.
This algorithm can be applied to the BE as well.
Finally in Section 6 we describe ongoing and future work.

\section{BE, RBE and associated SDEs in the laboratory frame}

In a semiclassical probabilistic description of an electron 
or positron bunch the
spin-orbit dynamics is described by the
{\it spin-$1/2$ Wigner function} $\rho$ 
(also called the {\it Stratonovich function})
written as
\begin{eqnarray}
 \rho(t,z) = \frac{1}{2} \left( f(t,z) I_{2\times 2}
+ \vec{\sigma}\cdot\vec{\eta}(t,z) \right),
               \label{eq:rho}
\end{eqnarray}
where $f$ is the classical phase-space density
normalized by $\int f(t,z)dz = 1$ and
$\vec{\eta}$ is the
polarization density of the bunch. 
Here $z=(\vec{r},\vec{p})$ where $\vec{r}$ and $\vec{p}$
are the position and momentum vectors
of the phase space and $t$ is the time.
Also, $\vec{\sigma}$ is the vector of the three Pauli matrices.
Thus $f=Tr[\rho]$ and
$\vec{\eta}=Tr[\rho\vec{\sigma}]$. Here and in the following we use arrows on 
three-component column vectors
and no arrows on other quantities. 
As explained in Ref. \citen{BH}, $\vec{\eta}$ is proportional to the
spin angular momentum density.
In fact it is given by $\vec{\eta}(t,z)=f(t,z)\vec{P}_{loc}(t,z)$
where $\vec{P}_{loc}$ is the
local polarization vector.
Then $\rho(t,z) $ is a product of $f(t,z)$ and a pure spin part with 
$\rho(t,z) = \frac{1}{2} f(t,z)( I_{2\times 2}
+ \vec{\sigma}\cdot\vec{P}_{loc}(t,z) )$.
The polarization vector $\vec{P}(t)$ of the bunch is
$\vec{P}(t) = \int \vec{\eta}(t,z)dz$. 
When the particle motion is governed just by a Hamiltonian, as
in the case of protons 
where one neglects all synchrotron radiation effects,
the phase-space density is conserved along a trajectory. Then, 
the
polarization density obeys the Thomas-BMT equation along each trajectory.
However, if the particles are subject to noise and damping due to
synchrotron radiation, the evolution of the density
of particles in phase space is more complicated. But as advertised above it
can be handled with a Fokker-Planck formalism.

Then, by neglecting collective effects and after several other approximations,
the phase-space density evolves according to Ref. \citen{DK75}
via
\ba
&& \partial_t f  =  L_{FP}(t,z)f \; .
\label{eq:n10a}
\ea
Using the units as in Ref. \citen{DK75}
the Fokker-Planck operator $L_{FP}$ is defined by
\ba
&&
\hspace{-10mm}
L_{FP}(t,z) := -\nabla_{\vec{r}}\cdot\frac{1}{m\gamma}\vec{p}
-\nabla_{\vec{p}}\cdot [e\vec{E}(t,\vec{r})+\frac{e}{m\gamma}(\vec{p}
\times\vec{B}(t,\vec{r}))
\nonumber\\
&&  \hspace{-15mm}
+\vec{F}_{rad}(t,z)+\vec{Q}_{rad}(t,z)]
+\frac{1}{2}\sum_{i,j=1}^3 \partial_{p_i}\partial_{p_j}
{\cal E}_{ij}(t,z) \; ,
\label{eq:4.4.4}
\ea
where
%
\ba
&&
\vec{F}_{rad}(t,z):
=-\frac{2}{3} \frac{e^4}{m^5\gamma}
|\vec{p}\times\vec{B}(t,\vec{r})|^2
\vec{p}\; ,
\label{eq:4.4.9b} \\
&& Q_{rad,i}(t,z):=\frac{55}{48\sqrt{3}}
\sum_{j=1}^3\;
\frac{\partial[\lambda(t,z)p_i p_j]}{\partial p_j} \; ,
\label{eq:4.4.11} \\
&&
{\cal E}_{ij}(t,z):=\frac{55}{24\sqrt{3}}\lambda(t,z)p_i p_j  \; , \quad
\lambda(t,z)
:= \hbar \frac{|e|^5}{m^8\gamma}
|\vec{p}\times\vec{B}(t,\vec{r})|^3 \; ,
\label{eq:4.4.9c} \\
&&
\gamma\equiv \gamma(\vec{p})
=\frac{1}{m}\sqrt{|\vec{p}|^2+m^2} \; ,
\label{eq:4.4.9d}
\end{eqnarray}
and with $e$ and $m$ being the charge 
and rest mass of the electron or positron
and $\vec{E},\vec{B}$
being the external electric and magnetic fields. 

The so-called parabolic Fokker-Planck terms
are those in the double sum of (\ref{eq:4.4.4}).
The Fokker-Planck operator $L_{FP}(t,z)$ whose explicit form is taken from 
Ref. \citen{DK75}
is a linear second-order partial
differential operator and, with some additional approximations, is
commonly used for electron synchrotrons and
storage rings, see Section 2.5.4 in Ref. \citen{CT}
and Ref. \citen{Sa}.
As usual, since it is minuscule compared to all other forces,
the Stern-Gerlach effect from the spin onto the
orbit is neglected in (\ref{eq:n10a}).
The polarization density $\vec{\eta}$ evolves via eq. 2 in 
Ref. \citen{DK75}, i.e., via the laboratory-frame BE
\ba
&& \partial_t\vec{\eta}  =
L_{\rm FP}(t,z)\vec{\eta}+ M(t,z)\vec{\eta}
\nonumber\\
&&\quad
-[1+\nabla_{\vec{p}}\cdot \vec{p}]
\lambda(t,z)\frac{1}{m\gamma}
\frac{\vec{p}\times
\vec{a}(t,z)}{|\vec{a}(t,z)|}f(t,z) \; ,
%
\label{eq:n10}
\ea
where
\ba
&& M(t,z):=\Omega(t,z)-\lambda(t,z)\frac{5\sqrt{3}}{8}[I_{3\times 3}
-\frac{2}{9m^2\gamma^2}\vec{p}\vec{p}^T] \; ,
\label{eq:4.4.17} 
\ea
and with
\ba
&& \vec{a}(t,z):=\frac{e}{m^2\gamma^2}(\vec{p}\times
\vec{B}(t,\vec{r})) \; .
\label{eq:4.4.17a} 
\ea
%
%
%
The skew-symmetric matrix $\Omega(t,z)$
takes into account the Thomas-BMT spin-orbit coupling and thereby the depolarization.
The quantum aspect of 
(\ref{eq:n10a}) and (\ref{eq:n10}) 
is embodied in the factor 
$\hbar$ in $\lambda(t,z)$. For example $\vec{Q}_{rad}$ is
a quantum correction to the classical radiation reaction force $\vec{F}_{rad}$.
The terms $-\lambda(t,z)\frac{5\sqrt{3}}{8}\vec{\eta}$
and $\lambda(t,z)\frac{1}{m\gamma}
\frac{\vec{p}\times\vec{a}(t,z)}{|\vec{a}(t,z)|}f(t,z)$
take into account spin flips due to synchrotron radiation
and encapsulate the Sokolov-Ternov effect.
The term $\lambda(t,z)\frac{5\sqrt{3}}{8}
\frac{2}{9m^2\gamma^2}\vec{p}\vec{p}^T\vec{\eta}$
encapsulates the Baier-Katkov correction, and the term
$\nabla_{\vec{p}}\cdot \vec{p}
\;\lambda(t,z)\frac{1}{m\gamma}
\frac{\vec{p}\times\vec{a}(t,z)}{|\vec{a}(t,z)|}f(t,z)
=\sum_1^3 \partial_{p_i} [p_i
\;\lambda(t,z)\frac{1}{m\gamma}
\frac{\vec{p}\times\vec{a}(t,z)}{|\vec{a}(t,z)|}f(t,z)]$
encapsulates the kinetic-polarization effect.

%
The Ito SDEs corresponding to (\ref{eq:n10a}) can be written informally as
\ba
&& \frac{d\vec{r}}{dt}= \frac{1}{m\gamma}\vec{p} \; ,
\label{eq:4.4.18a} \\
&& \frac{d\vec{p}}{dt}= e\vec{E}(t,\vec{r})+\frac{e}{m\gamma}
(\vec{p}\times\vec{B}(t,\vec{r}))
+ \vec{F}_{rad}(t,z)
+\vec{Q}_{rad}(t,z)
+ \vec{\cal B}^{orb}(t,z)\xi(t) \; ,
\label{eq:4.4.18b}
\ea
where $\xi$ is the white noise process and
\ba
&& \vec{\cal B}^{orb}(t,z):=\vec{p}
\sqrt{ \frac{55}{24\sqrt{3}}\lambda(t,z)}\; ,
\label{eq:4.4.20b}
\ea
or more concisely as
%
\ba
&& \frac{dZ}{dt}= F(t,Z) + G(t,Z)\xi(t) \; .
\label{eq:4.4.18e}
\ea
More precisely, the stochastic process $Z=(\vec{r},\vec{p})^T$ evolves according to the integral equation
\ba
&& Z(t) = Z(t_0) + \int_{t_0}^t\; F(\tau,Z(\tau))d\tau
+  \int_{t_0}^t\; G(\tau,Z(\tau))d{\cal W}(\tau) \; ,
\label{eq:4.4.18f}
\ea
where the second integral in (\ref{eq:4.4.18f}) is the so-called
Ito integral and ${\cal W}$ is the Wiener process.
Note that in (\ref{eq:4.4.18e}), and from now on, the dependent variables 
in the SDEs are denoted by large letters. In contrast, independent variables
are denoted by small letters, as in $f(t,z)$.
We note that (\ref{eq:4.4.18e}) is ambiguous. 
It is common to interpret (\ref{eq:4.4.18e}) as either
an Ito system of SDEs or a Stratonovich system of SDEs, leading 
to different Fokker-Planck equations if
$G$ depends on $z$. The SDEs 
(\ref{eq:4.4.18e}) lead to 
(\ref{eq:n10a}) via Ito but not via Stratonovich.
In this paper all SDEs are to be interpreted in the Ito sense.
Helpful discussions about Ito SDEs can be found, for example, in Refs.
\citen{Ar,Ga1,Ga2}.

A remarkable and perhaps unknown fact
is our recent finding that the BE
can be modeled
in terms of white noise as well, i.e., we can construct  a system of
SDEs underlying (\ref{eq:n10a}) and (\ref{eq:n10}). 
We already have  (\ref{eq:4.4.18e}) for the orbital motion 
and now introduce a vector $\vec{S}$ defined to obey
%
\ba
&& \hspace{-5mm}
\frac{d\vec{S}}{dt}=M(t,Z)\vec{S}
+ \vec{\cal D}^{spin}(t,Z)+ \vec{\cal B}^{kin}(t,Z)\xi(t) \; ,
\label{eq:4.4.18c}
\ea
%
where
\ba
&&  \vec{\cal D}^{spin}(t,z):=-\lambda(t,z)
\frac{1}{m\gamma}
\frac{\vec{p}\times\vec{a}(t,z)}{|\vec{a}(t,z)|}
\; ,
\label{eq:4.4.20a} \\
&& \vec{\cal B}^{kin}(t,z):=
\frac{1}{m\gamma}
\frac{\vec{p}\times\vec{a}(t,z)}{|\vec{a}(t,z)|}
\sqrt{ \frac{24\sqrt{3}}{55}\lambda(t,z)} \; .
\label{eq:4.4.20c}
\ea
The terms $M(t,Z)$, $\vec{\cal B}^{kin}(t,z)$ and $\vec{\cal D}^{spin}(t,z)$ in (\ref{eq:4.4.18c}) are chosen so that they deliver
the required BE (\ref{eq:n10}) by the end of the path for obtaining the FPE described below.
As can be expected  from the discussion after (\ref{eq:4.4.17}) above,
the term $\Omega(t,Z)\vec{S}$ will
account for the Thomas-BMT spin-precession effect,
the terms $-\lambda(t,Z)\frac{5\sqrt{3}}{8}\vec{S}$
and $\vec{\cal D}^{spin}(t,Z)$ will
account for spin flips due to synchrotron radiation
and encapsulate the Sokolov-Ternov effect. 
The term proportional to $2/9$ in (\ref{eq:4.4.17})
will account for  the Baier-Katkov correction, and the white-noise term
$\vec{\cal B}^{kin}(t,Z)\xi(t)$ will 
account for  the kinetic-polarization effect. The latter 
motivates the use of the superscript ``kin''.
As the notation suggests, the white-noise process $\xi(t)$ in
(\ref{eq:4.4.18c}) is the same as the white-noise process  
$\xi(t)$ in (\ref{eq:4.4.18b}).


To show that (\ref{eq:4.4.18e}) and (\ref{eq:4.4.18c}) 
lead to (\ref{eq:n10a}) and
(\ref{eq:n10}) one proceeds as follows.
The SDEs for the joint process $(Z,\vec{S})$ can be written as
\begin{eqnarray}
\frac{d}{dt} 
\left(\begin{array}{c}Z\\ \vec{S} \end{array}\right) 
= H(t,Z,\vec{S}) + N(t,Z)\xi(t)
\end{eqnarray}
where
\begin{eqnarray}
H(t,Z,\vec{S})=\left(\begin{array}{c}F(t,Z)\\ M(t,Z)\vec{S} + \vec{\cal D}^{spin}(t,Z)\end{array}\right), \;\;\;\;
N(t,Z)=\left(\begin{array}{c}G(t,Z)\\ \vec{\cal B}^{kin}(t,Z)\end{array}\right),
\end{eqnarray}
and we remind the reader that the SDE is to be interpreted as an Ito SDE.
The associated Fokker-Planck equation for the $(Z,\vec{S})$ process evolves the (joint) probability density ${\cal P}={\cal P}(t,z,\vec{s})$
which is related to $f$ and $\vec{\eta}$ via
\begin{eqnarray}
&& f(t,z)=\int_{{\mathbb R}^3}\; d\vec{s}
{\cal P}(t,z,\vec{s}) \; , \quad
\vec{\eta}(t,z)=\int_{{\mathbb R}^3}\; d\vec{s}
\vec{s}{\cal P}(t,z,\vec{s}) \; .
\label{eq:4.4.30}
\end{eqnarray}
It is straightforward to show via the Fokker-Planck equation for ${\cal P}$
that $f$ and $\vec{\eta}$ evolve according to (\ref{eq:n10a}) and
(\ref{eq:n10}). Thus indeed
(\ref{eq:4.4.18e}) and (\ref{eq:4.4.18c}) lead to
(\ref{eq:n10a}) and (\ref{eq:n10}).  

Note that $|\vec{S}(t)|$ in (\ref{eq:4.4.18c}) is not conserved in 
time. So $\vec{S}(t)$ in (\ref{eq:4.4.18c}) is not the spin vector of
a single particle. Nevertheless, $\vec{S}(t)$ can be related to familiar quantities.
In fact, by (\ref{eq:4.4.30}) and since $f$ is the
phase-space density, at time $t$ the conditional expectation 
of $\vec{S}(t)$ given $Z(t)$ is 
$\frac{1}{f(t,z)}\vec{\eta}(t,z)$, namely the local polarization
$\vec{P}_{loc}(t,Z(t))$.
%
%

Because $\vec{P}(t)= \int \vec{\eta}(t,z)dz$ it
also follows from (\ref{eq:4.4.30}), that the polarization vector
$\vec{P}(t)$ is the expectation value of the random vector
$\vec{S}(t)$, i.e., $\vec{P}(t)=<\vec{S}(t)>$ with $\vec{S}(t)$ from
(\ref{eq:4.4.18c}). Thus, and
since $|\vec{P}(t)|\leq 1$, we obtain $|<\vec{S}(t)>|\leq 1$. In particular
the constraint on the initial condition is: $|<\vec{S}(0)>|\leq 1$.

Since (\ref{eq:n10a}) and (\ref{eq:n10}) follow from 
(\ref{eq:4.4.18e}) and (\ref{eq:4.4.18c}) one can use
(\ref{eq:4.4.18e}) and (\ref{eq:4.4.18c}) as the basis
for a Monte-Carlo spin tracking algorithm for $\vec{P}(t)$. Thus 
this would extend the standard Monte-Carlo spin tracking algorithms 
by taking into account all physical effects described by (\ref{eq:n10}),
like the Sokolov-Ternov effect,
the Baier-Katkov correction, the kinetic-polarization effect and, of course,
spin diffusion. A detailed paper on this 
is in progress \cite{ABBEH}. 

If we ignore the spin flip terms and the kinetic-polarization term 
in the BE then
(\ref{eq:n10}) simplifies to 
%
\begin{eqnarray}
 \partial_t\vec{\eta} =  L_{FP}(t,z)\vec{\eta}
+ \Omega(t,z(t) )\vec{\eta} \; .
\label{eq:n11c}
\end{eqnarray}
We refer to (\ref{eq:n11c}) as the 
reduced Bloch equation (RBE).
Accordingly the system of SDEs underlying (\ref{eq:n11c}) is
(\ref{eq:4.4.18e}) and a simplified (\ref{eq:4.4.18c}), namely
\ba
&& \frac{d\vec{S}}{dt}=\Omega(t,Z(t))\vec{S} \; .
\label{eq:4.4.18d}
\ea
%
The RBE models spin diffusion due to the orbital motion.
Note that by (\ref{eq:4.4.18d}), and in contrast to (\ref{eq:4.4.18c}),
$|\vec{S}(t)|$ is conserved in time.
As mentioned in the Introduction, 
the RBE is sufficient for computing the depolarization time and 
it shares the terms with the BE that are most challenging to discretize.

The conventional Monte-Carlo spin tracking algorithms 
to compute the radiative depolarization time, e.g., 
SLICKTRACK by D.P. Barber, SITROS by J. Kewisch, Zgoubi by F. Meot,
PTC/FPP by E. Forest, and Bmad by D. Sagan
take care of the spin diffusion and they are based on or are closely related 
to the SDEs (\ref{eq:4.4.18e}) and (\ref{eq:4.4.18d})
\cite{CT,Zg,Fo,Sag}.
In contrast the Monte-Carlo spin tracking algorithm proposed above is
based on the SDEs (\ref{eq:4.4.18e}) and (\ref{eq:4.4.18c})
taking into account spin diffusion, the Sokolov-Ternov effect,
the Baier-Katkov correction, and the kinetic-polarization effect.

The equations (\ref{eq:n10a}) and (\ref{eq:n10}) can be derived
from quantum electrodynamics, using 
the semiclassical approximation of the
Foldy-Wouthuysen transformation of the Dirac Hamiltonian and finally by making
a Markov approximation \cite{He}.
We stress however, that
(\ref{eq:4.4.18e}) and (\ref{eq:4.4.18c}) provide a
model for (\ref{eq:n10}) which can be treated classically.
In fact, in the special case where one neglects
all spin flip effects and the kinetic-polarization effect
the corresponding SDEs
(\ref{eq:4.4.18e}) and (\ref{eq:4.4.18d})
(and thus the RBE (\ref{eq:n11c})) can be derived purely
classically as in Ref. \citen{BH}.
See Section 3 too.

\section{RBE and SDEs in the beam frame}

In the beam frame, i.e., in accelerator coordinates $y$,
the RBE (\ref{eq:n11c}) becomes
\ba
&& \partial_\theta\vec{\eta}_Y 
= L_Y(\theta,y)\vec{\eta}_Y + 
\Omega_Y(\theta,y)\vec{\eta}_Y \; ,
\label{eq:3.10}
\ea
where the meaning of the subscript ``Y'' will become clear below.
Here $\theta$ is the accelerator azimuth,
\ba
&& L_Y(\theta,y)   =
-\sum_{j=1}^6 \partial_{y_j}
\biggl({\cal A}(\theta)y\biggr)_j
+ \frac{1}{2}b_Y(\theta)\partial_{y_6}^2 \; ,
\nonumber
\ea
${\cal A}(\theta)$ is a $6\times 6$ matrix encapsulating radiationless
motion and the deterministic effects of synchrotron radiation,
$b_Y(\theta)$ encapsulates the quantum fluctuations,
and $\Omega_Y(\theta,y)$  encapsulates the Thomas-BMT term. The latter
is a skew-symmetric $3\times 3$ matrix 
and we linearize it as in Ref.
\citen{RR}.
Note that ${\cal A}(\theta),\;
\Omega_Y(\theta,y)$ and $b_Y(\theta)$ are
$2\pi$-periodic in $\theta$.
Given the beam-frame polarization density $\vec{\eta}_Y$, the
beam-frame polarization vector $\vec{P}(\theta)$ of the bunch at
azimuth $\theta$ is
\begin{eqnarray}
\vec{P}(\theta)
=\int dy\; \vec{\eta}_Y(\theta,y)
\label{eq:3.12}
\end{eqnarray}
Our central
computational focus is the RBE (\ref{eq:3.10})
with $\vec{P}(\theta)$ being a quantity of interest.
To proceed with this 
we use the underlying system of SDEs which are
%
\ba
&& Y' = {\cal A}(\theta)Y + \sqrt{b_Y(\theta)} e_6\xi(\theta) \; ,
\label{eq:3.14} \\
&& \vec{S}' = \Omega_Y(\theta,Y)\vec{S} \; ,
\label{eq:3.15}
\end{eqnarray}
where $\xi$ is the white noise process,
$e_6=(0,0,0,0,0,1)^T$.
The six components of $Y$ are defined here as in Refs. \citen{RR} and 
\citen{CT2}.
Thus the sixth component of $Y$ is $(\gamma-\gamma_r)/\gamma_r$
where $\gamma_r$ is the reference value of $\gamma$.
Since (\ref{eq:3.14}) is an Ito system of SDEs
which, in the language of SDEs,
is linear in the
narrow sense, it defines a Gaussian process $Y(t)$ if $Y(0)$ is Gaussian.
See Ref. \citen{Ga1}.
Eqs. (\ref{eq:3.14}) and (\ref{eq:3.15}) can
%
%
be obtained by transforming 
(\ref{eq:4.4.18e}) and (\ref{eq:4.4.18d})
%
%
%
from the laboratory frame to the beam frame.
However
(\ref{eq:3.14}) and (\ref{eq:3.15}) 
can also be found in 
several expositions on spin in
high-energy electron storage rings, e.g., Ref. \citen{RR}.
Note that these expositions make some approximations.
We use Ref. \citen{RR} which involves transforming from the laboratory to the beam frame
and then linearizing in the beam-frame coordinates, leading to the linear
SDEs (\ref{eq:3.14}) and to $\Omega_Y(\theta,Y)$ which is linear in $Y$. 
Practical calculations with the Derbenev-Kondratenko formalism make similar
approximations. 

The Fokker-Planck equation for the density of the Gaussian process $Y$ is
\ba
&& \partial_\theta {\cal P}_Y = L_Y(\theta,y) \ {\cal P}_Y \; .
\label{eq:3.18}
\end{eqnarray}
%
In fact with
(\ref{eq:3.14}) and (\ref{eq:3.15}) the evolution equation for
the spin-orbit joint probability density ${\cal P}_{YS}$
is the following Fokker-Planck equation
\ba
&& \hspace{-10mm}
\partial_\theta {\cal P}_{YS} = L_Y(\theta,y){\cal P}_{YS}
-\sum_{j=1}^3 \partial_{s_j}
\Biggl(\biggl(\Omega_Y(\theta,y)\vec{s}\biggr)_j  {\cal P}_{YS}\Biggr) \; .
\label{eq:3.20}
\end{eqnarray}
Note that ${\cal P}_Y$ is related to ${\cal P}_{YS}$ by
\begin{eqnarray}
{\cal P}_Y(\theta,y)=
\int_{{\mathbb R}^3}\; d\vec{s} \;{\cal P}_{YS}(\theta,y,\vec{s}) \; .
\label{eq:3.21}
\end{eqnarray}
%
Also, by integrating (\ref{eq:3.20}) over $\vec{s}$ one 
recovers (\ref{eq:3.18}).
The polarization density $\vec{\eta}_Y$ corresponding
to ${\cal P}_{YS}$ is defined by
\begin{eqnarray}
\vec{\eta}_Y(\theta,y)=
\int_{{\mathbb R}^3}\; d\vec{s} \;
\vec{s}\;{\cal P}_{YS}(\theta,y,\vec{s}) \; .
\label{eq:3.22}
\end{eqnarray}
Note that (\ref{eq:3.21}) and (\ref{eq:3.22}) are analogous to
(\ref{eq:4.4.30}).
The RBE (\ref{eq:3.10}) follows from
(\ref{eq:3.20}) by differentiating (\ref{eq:3.22}) w.r.t. $\theta$.
For (\ref{eq:3.10}) see Ref. \citen{BH} too.
We recall that the relation between a system of
SDEs and its Fokker-Planck equation is standard, see, e.g.,
Refs. \citen{Ar,Ga1,Ga2}.
%

\section{Approximating the beam-frame
RBE by the method of averaging}

Because the coefficients of $L_Y(\theta,y)$ are $\theta$-dependent, the
RBE (\ref{eq:3.10}) is difficult to understand analytically and difficult for a numerical method.
Since the RBE is derivable from the associated SDEs (\ref{eq:3.14}) and (\ref{eq:3.15}) we can focus on these difficulties 
in the SDEs, rather than in the RBE,
where approximation methods are better developed.
For this purpose we rewrite (\ref{eq:3.14}) as
\ba
&& Y' = (A(\theta)+\epsilon \delta A(\theta))Y
+ \sqrt{\epsilon}\sqrt{b(\theta)}
e_6\xi(\theta)
\label{eq:4.10}
\end{eqnarray}
where $A(\theta)$ is the Hamiltonian part of $\cal A(\theta)$
and $\epsilon$ is chosen so that $\delta A$ is order 1.
Then $b$ is defined by $\sqrt{\epsilon}\sqrt{b(\theta)}=\sqrt{b_Y(\theta)}$.
Here  $\epsilon\delta A(\theta)$ represents the part of
$\cal A(\theta)$ associated with damping effects due to
synchrotron radiation and cavities (see, e.g., eq. 5.3 in
Ref. \citen{RR}).
The term $\sqrt{\epsilon}\sqrt{b(\theta)}$ corresponds to the quantum noise and
the square root is needed for the balance of damping, cavity acceleration and quantum noise
(See Eq. (\ref{eq:4.16})).
We are interested in situations where $Y$ has been appropriately
scaled and where the synchrotron radiation has a small effect so that $\epsilon$ is small.

Eq. \ref{eq:4.10} can be approximated using the method of averaging which will eliminate
some of the $\theta$ dependent coefficients and allow for a numerical method which can integrate the 
resultant RBE efficiently over long times.
This has the added benefit of deepening our analytical understanding,
as a perturbation analysis usually does.
We call the approximation of the RBE the effective RBE and we will
find it by refining the averaging
technique presented 
in Section 2.1.4
of the Accelerator Handbook \cite{CT2}.
This refinement allows us to use the method of averaging
to approximate the SDEs
(\ref{eq:4.10}). 
We just give a sketch here (a detailed account will be published elsewhere \cite{El}).
%
%
%
%

Because the process $Y$ is Gaussian, if $Y(0)$ is Gaussian,
all the information is in its
mean $m_Y$ and covariance $K_Y$ and they evolve by the ODEs
\ba
&& m_Y' = (A(\theta)+\epsilon \delta A(\theta))m_Y \; ,
\label{eq:4.15} \\
&& K_Y' = (A(\theta)+\epsilon \delta A(\theta))K_Y+K_Y(A(\theta)
+\epsilon \delta A(\theta))^T
+\epsilon b(\theta) e_6^{} e_6^T
\label{eq:4.16}
\end{eqnarray}
In (\ref{eq:4.16}) the $\delta A$ terms and the b  are
balanced at $O(\epsilon)$ and so can be treated together in first order
perturbation theory. This is the reason for the $\sqrt{\epsilon}$ in
(\ref{eq:4.10}).
However this balance is also physical since the damping and diffusion come from the same source
and the cavities replenish the energy loss.
We cannot include the spin equation
(\ref{eq:3.15}) because the joint $(Y,\vec{S})$ process is not
Gaussian. Eq. (\ref{eq:3.15}) has a quadratic nonlinearity 
since it is linear in $Y$ and $\vec{S}$ so that the
joint moment equations would not close. Thus here we will 
apply averaging to the $Y$ process only and discuss the spin after that.
However, see Remark 3 below which outlines a plan for a combined approach.

To apply the method of averaging to (\ref{eq:4.15}) and (\ref{eq:4.16})
we must transform them to a standard form for averaging. We do this
by using a fundamental solution matrix $X$ of the
unperturbed $\epsilon=0$ part of (\ref{eq:4.10}) and (\ref{eq:4.15}), i.e.,
%
\ba
&&X'=A(\theta)X \; .
\label{eq:4.18}
\end{eqnarray}
We thus transform $Y$, $m_Y$ and $K_Y$ into $U$, $m_U$ and $K_U$ via
\ba
Y=X(\theta) U , \quad  m_Y=X(\theta)m_U , \quad  K_Y=X(\theta)K_UX^T(\theta)
\label{eq:4.20}
\end{eqnarray}
and (\ref{eq:4.10}), (\ref{eq:4.15}) and (\ref{eq:4.16}) are transformed to
\ba
&& U' = \epsilon {\cal D}(\theta)U
+ \sqrt{\epsilon}\sqrt{b(\theta)}
X^{-1}(\theta)e_6\xi(\theta)
\label{eq:4.31} \\
&& m_U' = \epsilon {\cal D}(\theta)m_U \; ,
\label{eq:4.25} \\
&& K_U' = \epsilon ({\cal D}(\theta)K_U +K_U {\cal D}^T(\theta))
+ \epsilon {\cal E}(\theta)
\label{eq:4.26}
\end{eqnarray}
Here ${\cal D}(\theta)$ and ${\cal E}(\theta)$ are defined by
\ba
&& {\cal D}(\theta)=X^{-1}(\theta)\delta A(\theta)X(\theta) \; ,
\label{eq:4.28} \\
&& {\cal E}(\theta)
= b(\theta)X^{-1}(\theta)e_6^{} e_6^TX^{-T}(\theta)
\label{eq:4.29}
\end{eqnarray}
Of course,  (\ref{eq:4.31}) - (\ref{eq:4.26}) carry the same information as (\ref{eq:4.10}) - (\ref{eq:4.16}).

Now, applying the method of averaging to (\ref{eq:4.25}) and (\ref{eq:4.26}), we obtain the Gaussian process $V$ with mean and covariance matrix
\ba
&& m_V' = \epsilon \bar{\cal D}m_V \; ,
\label{eq:4.35} \\
&& K_V' = \epsilon ( \bar{\cal D}K_V +K_V \bar{\cal D}^T )+
\epsilon \bar{\cal E}
\label{eq:4.36}
\end{eqnarray}
where the bar denotes $\theta$-averaging, i.e., the operation
$\lim_{T\rightarrow\infty}(1/T)\int_0^T d\theta\cdots$.
For physically reasonable $A$ each fundamental matrix $X$
is a quasiperiodic function whence
${\cal D}$ and ${\cal E}$ are quasiperiodic functions so that
their $\theta$  averages $\bar{\cal D}$ and $\bar{\cal E}$ exist.
By averaging theory $|m_U(\theta)-m_V(\theta)|\le C_1(T)\epsilon$
and  $|K_U(\theta)-K_V(\theta)| \le C_2(T) \epsilon$ for $0\leq \theta\leq T/\epsilon$
where $T$ is a constant (see also Refs.
\citen{SSC,EHVG,SVM,Mu}) and $\epsilon$ small.
However, we expect to be able to show that these estimates are uniformly 
valid on $[0,\infty)$  so that an accurate estimate of
the orbital equilibrium would be found.

The key point now is that every Gaussian process $V$, whose mean
$m_V$ and covariance matrix $K_V$
satisfy the ODEs (\ref{eq:4.35}) and (\ref{eq:4.36}), also satisfies
the system of SDEs
\ba
&& V' = \epsilon \bar{\cal D}V + \sqrt{\epsilon}{\cal B}
(\xi_1,...,\xi_k)^T \; .
\label{eq:4.40}
\end{eqnarray}
Here $\xi_1,...,\xi_k$ are statistically independent versions of
the white noise process and
${\cal B}$ is a $6\times k$ matrix which satisfies
${\cal B}{\cal B}^T=\bar{\cal E}$
with $k=rank(\bar{\cal E})$.
Since $m_U(\theta)=m_V(\theta)+O(\epsilon)$
and  $K_U(\theta)=K_V(\theta)+O(\epsilon)$ we get $U(\theta)\approx V(\theta)$. In particular
$Y(\theta)\approx
X(\theta)V(\theta)$ (more details will be in 
Ref. \citen{El}).
Conversely, the mean vector $m_V$ and covariance matrix $K_V$ of
every $V$ in (\ref{eq:4.40})
satisfy the ODEs (\ref{eq:4.35}) and (\ref{eq:4.36}).

\noindent {\bf Remark}: \\
It's likely that stochastic averaging techniques 
can be applied directly to (\ref{eq:4.31})
giving (\ref{eq:4.40}) as an approximation and we are looking into this
(see Ref. \citen{CE} and references therein). 
However, because (\ref{eq:4.31}) is linear and defines a Gaussian process, the theory for getting to (\ref{eq:4.40})
from the ODEs for the moments could not be simpler, even though it is indirect.
\hfill $\Box$

To proceed with an analysis of (\ref{eq:4.40}) and its associated 
Fokker-Planck equation
we need an appropriate $X$ and we note that
$X(\theta)=M(\theta)C$
where $C$ is an arbitrary invertible $6\times 6$ matrix and
$M$ is the principal solution matrix, i.e.,
$M'=A(\theta)M,M(0)=I$. Thus choosing $X$ boils down to
choosing a good $C$.
As is common for spin physics in electron storage rings
we emulate Chao's approach 
(see Section 2.1.4 in Ref.
\citen{CT2} and Refs. \citen{Ch1,Ch2}) and use the
eigenvectors of $M(2\pi)$.
We assume that the unperturbed orbital motion is stable.
Thus $M(2\pi)$ has a full set of linearly independent eigenvectors
and the eigenvalues are on the unit circle in the complex
plane \cite{MHO}.
We further assume a non-resonant condition on the orbital frequencies.
We construct $C$ as a real matrix using the real and
imaginary parts of the
eigenvectors in its columns and using the fact that
$M(2\pi)$ is symplectic (since
$A(\theta)$ is a Hamiltonian matrix).
It follows that
$\bar{\cal D}$ has block diagonal
form and $\bar{\cal E}$ has diagonal form. 
Explicitly,
\begin{eqnarray}
&& \bar{\cal D}=
\left( \begin{array}{ccc} {\cal D}_I & 0_{2\times 2} & 0_{2\times 2} \\
0_{2\times 2} & {\cal D}_{II} & 0_{2\times 2} \\
0_{2\times 2} & 0_{2\times 2} & {\cal D}_{III} \end{array}\right) \; ,
\label{eq:5.15} \\
&& {\cal D}_\alpha = \left( \begin{array}{cc} a_\alpha & b_\alpha \\
-b_\alpha & a_\alpha \end{array}\right) \; , (\alpha=I,II,III)
\label{eq:5.16}
\end{eqnarray}
and $\bar{\cal E}=diag({\cal E}_I,{\cal E}_I,{\cal E}_{II},{\cal E}_{II},{\cal E}_{III},{\cal E}_{III})$ with $a_\alpha\leq 0$ and
${\cal E}_I,{\cal E}_{II},{\cal E}_{III}\geq 0$.

To include the spin note that,
under the transformation $Y\mapsto U$,
(\ref{eq:3.14}) and (\ref{eq:3.15}) become
\ba
&& U' = \epsilon {\cal D}(\theta)U
+ \sqrt{\epsilon}\sqrt{b(\theta)}
X^{-1}(\theta)e_6\xi(\theta)
\label{eq:4.31a} \\
&& \vec{S}' = \Omega_Y(\theta,X(\theta)U)\vec{S} \; ,
\label{eq:3.15a}
\end{eqnarray}
where we have repeated (\ref{eq:4.31}).
Now, as we just mentioned, 
$U$ is well approximated by $V$, i.e., $U=V+O(\epsilon)$ on
$\theta$ intervals of a length of $O(1/\epsilon)$ (and because of damping
we may have uniform validity for $0\leq \theta<\infty$). Thus
\ba
&& \Omega_Y(\theta,X(\theta)U)
= \Omega_Y(\theta,X(\theta)V)+O(\epsilon)
\label{eq:3.15b}
\end{eqnarray}
and (\ref{eq:3.15a}) becomes
\ba
&& \vec{S}' = \Omega_Y(\theta,X(\theta)V)\vec{S} +O(\epsilon) \; .
\label{eq:3.15ba}
\end{eqnarray}
Dropping the $O(\epsilon)$ in (\ref{eq:3.15ba}) and replacing
$U$ by $V$ in (\ref{eq:3.15a}) we obtain the system
\ba
&& V' = \epsilon \bar{\cal D}V + \sqrt{\epsilon}{\cal B}
(\xi_1,...,\xi_k)^T \; ,
\label{eq:4.45} \\
&& \vec{S}' = \Omega_Y(\theta,X(\theta)V)\vec{S} \; ,
\label{eq:4.46}
\end{eqnarray}
where (\ref{eq:4.45}) is a repeat of (\ref{eq:4.40}).
With (\ref{eq:4.45}) and (\ref{eq:4.46}) the evolution equation for
the spin-orbit probability density
${\cal P}_{VS}={\cal P}_{VS}(\theta,{\rm v},\vec{s})$
is the following Fokker-Planck equation:
\ba
&&  \hspace{-10mm}
\partial_\theta {\cal P}_{VS} = L_V(v){\cal P}_{VS}
-\sum_{j=1}^3 \partial_{s_j}
\Biggl(\biggl(\Omega_Y(\theta,X(\theta){\rm v})\vec{s}\biggr)_j  {\cal P}_{VS}\Biggr) \; ,
\label{eq:4.50}
\end{eqnarray}
where
\ba
&& L_V(v)
= -\epsilon\sum_{j=1}^6 \partial_{{\rm v}_j}
(\bar{\cal D} {\rm v})_j
+ \frac{\epsilon}{2}\sum_{i,j=1}^6 \bar{\cal E}_{ij}
\partial_{{\rm v}_i}\partial_{{\rm v}_j} \; .
\label{eq:4.51}
\end{eqnarray}
Thus the three degrees of freedom are uncoupled in $L_V$
since, by (\ref{eq:4.51}),
\ba
&& L_V = L_{V,I} + L_{V,II}+L_{V,III}
\label{eq:5.20}
\end{eqnarray}
where each $L_{V,\alpha}$ is an operator in one degree of freedom
(=two dimensions)
and is determined by ${\cal D}_\alpha$ and ${\cal E}_\alpha$
via (\ref{eq:4.51}) ($\alpha=I,II,III$).
This is important for our numerical approach.

The polarization density $\vec{\eta}_V$ corresponding
to ${\cal P}_{VS}$ is defined by
\begin{eqnarray}
\vec{\eta}_V(\theta,{\rm {\rm v}})
=\int_{{\mathbb R}^3}\; d\vec{s}
\vec{s}\;{\cal P}_{VS}(\theta,{\rm v},\vec{s})
\label{eq:4.52}
\end{eqnarray}
so that  by (\ref{eq:4.50}), the effective RBE is
\ba
&& \partial_\theta\vec{\eta}_V 
= L_V(v)\vec{\eta}_V + \Omega_Y(\theta,X(\theta){\rm v})\vec{\eta}_V \; .
\label{eq:4.53}
\ea
%
%
The coefficients of $L_V(v)$ are $\theta$-independent
for every choice of $X$ and this is necessary for our numerical
method.

We now have $Y(\theta)=X(\theta)U(\theta)\approx Y_a(\theta):=X(\theta)V(\theta)$
and it follows that $\vec{\eta}_Y$ in (\ref{eq:3.10}) is given approximately by
\begin{eqnarray}
\vec{\eta}_{Y}(\theta,y) \approx \vec{\eta}_{Y,a}(\theta,y) = \det(X^{-1}(0))
\vec{\eta}_V(\theta,X^{-1}(\theta){\rm y}) \; .
\label{eq:4.55}
\end{eqnarray}
Now (\ref{eq:4.53}) and the effective RBE for $\vec{\eta}_{Y,a}$
carry the same information.
However in general the effective 
RBE for $\vec{\eta}_{Y,a}$ does not have the
nice feature of (\ref{eq:4.53}), of being $\theta$-independent,
which make the latter useful for
our numerical method (see below). Hence
we discretize
(\ref{eq:4.53}) rather than the effective RBE for $\vec{\eta}_{Y,a}$.

We now make several remarks on the validity of the approximation
leading to (\ref{eq:4.45}) and (\ref{eq:4.46}) and thus to (\ref{eq:4.53}).\\

\noindent {\bf Remark 1}: \\
The averaging which leads to (\ref{eq:4.53}) affects only the orbital variables.
It was justified by using the fact that (\ref{eq:4.31a})
is linear whence it defines a
Gaussian process when the initial condition is Gaussian. This allowed us to 
apply the method of averaging 
to the first and second moments rather than the
SDEs themselves.\hfill $\Box$
\\
\noindent {\bf Remark 2}: \\
We cannot extend the moment approach to the system
(\ref{eq:4.31a}) and (\ref{eq:3.15a})
because (\ref{eq:3.15a}) has a quadratic
nonlinearity and the system of moment equations do not close.
In future work, we will pursue approximating the system
(\ref{eq:4.31a}) and (\ref{eq:3.15a}) using stochastic
averaging as in Ref. \citen{CE}. 
\hfill $\Box$ \\
\noindent {\bf Remark 3}: \\
Because of the $O(\epsilon)$ error in (\ref{eq:3.15ba}) we apriori
expect an error of $O(\epsilon \theta)$ in $\vec S$  when going from
(\ref{eq:3.15a}) to (\ref{eq:4.46})
and so (\ref{eq:4.53}) may only give a good approximation
to $\vec{\eta}_Y$ on $\theta$ intervals of a length of $O(1)$.
The work mentioned in Remark 2 above may shed
light on this. 
In addition we will split $\Omega_Y$ into two pieces:
$\Omega_Y(\theta,y) = \Omega_0(\theta)
+\epsilon_s \omega(\theta,y)$ 
where $\Omega_0$ is the closed-orbit contribution to $\Omega_Y$
and $\epsilon_s$ is chosen so that $\omega$ is $O(1)$.
Then, in the case where $\epsilon_s=\epsilon$, (\ref{eq:3.15a}) becomes
$\vec{S}' = \Omega_0(\theta) \vec{S}+\epsilon \omega(\theta,X(\theta)U)\vec{S}$.
By letting $\vec{S}(\theta)=\Psi(\theta)\vec{T}(\theta)$
where $\Psi'=\Omega_0(\theta)\Psi$ we obtain
\ba
&& \vec{T}'=\epsilon{\mathfrak D}(\theta,U)\vec{T} \; ,
\label{eq:3.15e}
\end{eqnarray}
where ${\mathfrak D}(\theta,U)
=\Psi^{-1}(\theta)\omega(\theta,X(\theta)U)\Psi(\theta)$.
Our system is now (\ref{eq:4.31a}) and (\ref{eq:3.15e}) and the associated
averaged system consists of (\ref{eq:4.45}) and of the averaged form of
(\ref{eq:3.15e}), i.e.,
\ba
&& V' = 
\epsilon \bar{\cal D}V + \sqrt{\epsilon}{\cal B}
(\xi_1,...,\xi_k)^T \; ,
\label{eq:3.15f}\\
&& \vec T_a'=\epsilon\bar{\mathfrak D}(V)\vec{T}_a \; .
\label{eq:3.15g}
\end{eqnarray}
It seems likely that
$\vec{S}(\theta)=\Psi(\theta)\vec{T}_a(\theta)+O(\epsilon)$
for $0\leq \theta<O(1/\epsilon)$, which we hope to prove.
\hfill $\Box$\\
\noindent {\bf Remark 4}: \\
We have applied the method of averaging to
a $1$-degree-of-freedom model (=2 dimensions)
with just one spin variable and have 
verified the $O(\epsilon)$ error
analytically. 
In addition, we are working on a 
$2$-degree-of-freedom model (=4 dimensions) 
with just one spin variable. These are
discussed in our two ICAP18 papers \cite{HABBE,OB}. 
These models will be helpful
for our $3$-degree-of-freedom study we outlined here.
\hfill $\Box$

\section{Sketch of the numerical approach}

We now briefly sketch our numerical approach to the
effective RBE (\ref{eq:4.53}). For more details see Ref. \citen{OB}.
The numerical computations are performed by using
3 pairs $(r_\alpha,\varphi_\alpha)$
of polar coordinates, i.e.,
${\rm v}_1=r_I \cos \varphi_I,...,{\rm v}_6=r_{III}\sin \varphi_{III}$.
The angle variables are Fourier transformed whence
the Fourier coefficients are functions of time and the radial variables.
We discretize the radial variables by using the
{\it collocation} method \cite{CHQZ,Forn}
using a Chebychev grid for each radial variable.
For each Fourier mode this results in
a system of linear first-order ODEs in  $\theta$
which we discretize by using an implicit/explicit
$\theta$-stepping scheme. 
The collocation method is a minimial-residue method
by which the residual of the PDE is zero at the numerical grid points.
Because of (\ref{eq:5.15}),(\ref{eq:5.16}) and (\ref{eq:4.51}), 
the Fourier modes are uncoupled
in $L_V\vec{\eta}_V$ so that the only coupling of Fourier modes in
(\ref{eq:4.53}) comes via 
$\Omega_Y(\theta,X(\theta){\rm v})\vec{\eta}_V$
and this
coupling is local since $\Omega_Y(\theta,X(\theta){\rm v})$ is linear in
${\rm v}$. Thus  the parabolic terms are separated from
the mode coupling terms. Hence
in the $\theta$ stepping $L_V\vec{\eta}_V$ is treated
implicitly and $\Omega_Y(\theta,X(\theta){\rm v})\vec{\eta}_V$ 
is treated explicitly.
We exploit the decoupling by evolving the resulting ODE system
with the additive Runge-Kutta (ARK) method. As described in Ref. \citen{ark}, ARK methods are high-order
semi-implicit
methods that are constructed from a set of consistent Runge-Kutta (RK) methods.
In the RBE the parabolic part of the equation is treated with a diagonally implicit RK method (DIRK)
and the mode coupling part is treated with an explicit RK (ERK) method
which does not require a linear solve.
The ODE system can be evolved independently in time for each Fourier mode,
resulting in a computational cost for each timestep that scales as $\mathcal{O}(N^{3q})$ per mode where $N$ is the number of grid-points for each of the
six dimensions and where 
$1 \le q \le 3$, depending on the algorithms used for the linear solve.
However, only algorithms with $q \approx 1$ are feasible 
(for Gaussian elimination $q=3$).
Fortunately, the structure of the averaged equations (e.g.,
the parabolic terms are decoupled from mode coupling  terms) 
allows efficient parallel implementation. 
We have applied this in a 
$1$-degree-of-freedom model 
and have demonstrated the spectral convergence.\cite{OB}

\section{Discussion and next steps}

We are continuing our work on the second model, i.e., the one based on the
Bloch-equation, by
extending the averaging and numerical work from the RBE to the BE and 
from one and two degrees of freedom to
three degrees of freedom, aiming towards realistic FODO lattices \cite{El,B}.
This will include depolarization and polarization times and equilibrium
polarization. 
Extending the second model from the RBE to the BE involves averaging
and thus involves the SDEs from the third model. Moreover we plan to use
the third model to develop a
Monte-Carlo spin tracking algorithm which is based on the SDEs
(\ref{eq:4.4.18e}) and (\ref{eq:4.4.18c}) and which
takes into account the Sokolov-Ternov effect,
the Baier-Katkov correction, the kinetic-polarization effect and
spin diffusion.
Furthermore we continue our work on comparing
the Bloch-equation approach with 
Derbenev-Kondratenko-formula approach and estimating the
polarization at the FCC-ee and CEPC.

\section{ Acknowledgement}
This material is based upon work supported by the U.S. Department
of Energy, Office of Science, Office of High Energy Physics, under Award
Number DE-SC0018008.


\end{document}